\documentclass[preprint,12pt]{elsarticle}
\usepackage{bm}
\usepackage{graphicx}

\newcommand{\figurewidth}{4.0in}
\newcommand{\subfigurewidth}{2.5in}
\newcommand{\subboxwidth}{2.7in}
\newcommand{\bu}         {{\mathbf u}}
\newcommand{\bv}         {{\mathbf v}}
\newcommand{\bw}         {{\mathbf w}}
\newcommand{\br}         {{\mathbf r}}

\journal{Computers \& Mathematics with Applications}

\begin{document}

\sloppy

\begin{frontmatter}

  \title{Multiscale modelling strategy using the \\
         lattice Boltzmann method for polymer dynamics \\
         in a turbulent flow}

  \author[MPIP]{Jonghoon Lee\corref{cor1}}
  \ead{jonglee@mpip-mainz.mpg.de} \cortext[cor1]{Corresponding
  author}
  \author[MPIP]{Burkhard D\"{u}nweg}
  \author[TUI]{J\"{o}rg Schumacher}

  \address[MPIP]{Max Planck Institute for Polymer Research, 
                 Ackermannweg 10, \\
                 D-55128 Mainz, Germany}
  \address[TUI]{Department of Mechanical Engineering, 
                Technische Universit\"{a}t Ilmenau, \\ 
                D-98684 Ilmenau, Germany}

  \date{\today}

  \begin{abstract}
    Polymer dynamics in a turbulent flow is a problem spanning several
    orders of magnitude of length and time scales. A microscopic
    simulation covering all those scales from the polymer segment to
    the inertial scale of turbulence seems impossible within the
    foreseeable future. We propose a multiscale simulation strategy to
    enhance the spatio-temporal resolution of local Lagrangian
    turbulent flow by matching two different simulation techniques,
    i.~e. direct numerical simulation for the flow as a whole, and the
    lattice Boltzmann method coupled to polymer dynamics at the
    Kolmogorov dissipation scale. Local turbulent flows sampled by
    Lagrangian tracer particles in the direct numerical simulation are
    reproduced in the lattice Boltzmann model with a finer resolution,
    by supplying the latter with both the correct initial condition
    as well as the correct time-dependent boundary condition, sampled
    from the former. When combined with a Molecular Dynamics simulation of
    a polymer chain in the lattice Boltzmann model, it provides a
    strategy to simulate the \emph{passive} dynamics of a polymer
    chain in a turbulent flow covering all those scales. While this
    approach allows for a fairly realistic model of the macromolecule,
    the back-coupling to the flow on the large scales is missing.
  \end{abstract}

\end{frontmatter}

\section{Introduction}

By adding long chain molecules in a concentration as minute as
$10^{-5}$ in weight, the frictional drag in a turbulent flow through a
pipe can be significantly reduced~\cite{Lum:1969}. This phenomenon of
drag reduction in turbulent flow by polymer additives has been known
for more than half a century~\cite{Tom:1949}, and numerous
experimental and theoretical studies have been
executed~\cite{Nad:1995}. However, it is widely acknowledged that the
theoretical understanding of the mechanism behind this phenomenon has
not yet gone beyond the tentative
level~\cite{Lum:1973,Tab:1986,Pro:2008}, and is under current debate.
Amongst others, not even the question whether drag reduction exists in
homogeneous turbulence~\cite{Sre:2000,Ben:2004} has been conclusively
resolved yet.

It is obvious that the flexibility of the polymer chains must play a
crucial role for this phenomenon, given the fact that the addition of
non-flexible colloidal particles does not result in such a dramatic
effect at such a dilution. In thermal equilibrium, a polymer chain
assumes a random coiled shape, and a convenient measure of its size is
the root mean square radius of gyration $R_g$~\cite{doi:1986}. In a
strong flow, however, a polymer chain may be stretched up to its
contour length, which, under such conditions, becomes the
characteristic size of the macromolecule. This can easily be orders of
magnitude larger than $R_g$ for a chain with a large degree of
polymerization. In the stretched state, the elongational viscosity
increases by factors of $\sim 10^{4}$ for large
molecules~\cite{Hin:1977}. An important attempt to understand the
turbulent drag reduction in pipe flow based on this coil-stretch
transition~\cite{deGennes:1974} and corresponding viscosity variation
was made by Lumley~\cite{Lum:1973}. Among other theories for drag
reduction, the argument by Tabor and de Gennes is based on the
entropic elasticity of polymer chains~\cite{Tab:1986}. They thought
that the coil-stretch transition would not occur under randomly
fluctuating strains, and the moderately stretched state would prevail,
which would not modify the viscosity drastically. They argued that the
elastic energy stored within the molecule would interfere with the
cascading turbulent energy at some length scale larger than the
Kolmogorov dissipation scale ($\eta$), where the crossover from
turbulent to laminar behavior occurs. So far, however, the
experimental difficulties in observing details of molecular motion in
a turbulent flow have prevented us from obtaining a decisive and
direct evidence for existing theories and hypotheses.

The main difficulty in the theoretical, experimental or computational
analysis of the drag reduction problem is the strong disparity of the
relevant length scales. The energy injection scale of the largest eddy
is usually orders of magnitude larger than the energy dissipation
Kolmogorov scale ($\eta$) of the smallest eddy. In a turbulent flow
with polymer additives, non-linear interactions strongly couple
different scales by transferring excitations from the injection scale
all the way down to $\eta$, which is coupled with the scales of
polymers. In a fully developed homogeneous turbulence, $\eta$
corresponds to $\sim L/{Re}^{\frac{3}{4}}$, where $L$ is the external
scale of flow and ${Re}$ is the Reynolds number. A textbook
Kolmogorovian turbulent flow with clear scale separations among
relevant scales requires a relatively large $Re$, which renders
negligible $\eta/L$. Even in state-of-the-art Direct Numerical
Simulations (DNS) of turbulent flow in a three-dimensional periodic
simulation box of $1024^3$ grid points using supercomputers, $\eta$
becomes as long as only a few grid spacings. Considering that $R_g$ in
typical experiments is about two orders of magnitude smaller than
$\eta$, even for systems with very high molecular weight and large
Reynolds numbers \cite{bodenschatz}, it becomes obvious that the
details of the polymer conformational degrees of freedom are
completely out of the available resolution of DNS. In other words, a
first-principles simulation, which covers all these length scales down
to the monomer length scale, and, simultaneously, all the
corresponding time scales, would need such a fine resolution that it
is unfeasible, even on the most advanced supercomputers.

Existing computational studies regarding polymer-flow interaction have
relied on simplification either of the turbulence model or of the
polymer model, or of both. One approach covers the scale around and
above the Kolmogorov scale. The turbulent flow can be simulated with
realistic coupling between eddies, but the polymer models used are as
simple as variations of the Hookean dumbbell model, which gives only
an extremely crude picture of the intricate interplay between the flow
and the molecule, whose frictional properties and relaxation time
spectrum depend very strongly on the
conformation~\cite{Afo:2005,Dav:2006,Cel:2006,Ter:2004}. The other
approach focuses on the purely dissipative sub-Kolmogorov scale.
Realistic chain models for the polymers can be used, but the flow
environment is over-simplified without interaction with larger
scales~\cite{Hur:2000,Sto:2003,Jen:2004}. A numerical simulation of
such dynamics with a realistic polymer model and a realistic turbulent
flow has so far not yet been done. Hence it is generally acknowledged
that a multiscale approach is necessary.

In this paper, we present such a multiscale simulation strategy to
simulate the dynamics of a linear polymer chain in a homogeneous
turbulent flow. Essentially, we achieve feasibility by splitting the
problem up into two stages that correspond to the large and small
scales, respectively. This comes at the price of losing some realism
concerning the consistent coupling of the scales: There is only a
``top-down'' coupling from the large to the small scales, but no
corresponding ``bottom-up'' back-coupling from the latter to the
former.

The large scales are therefore handled by just running a standard DNS
of turbulent flow, down to the Kolmogorov scale, by a
spectral code. The idea is that this is probably the best available
source of information about the fluctuating flow field which the
polymer chains experience on a finer scale. Certainly it is more
realistic than just a simple shear or elongational flow, or a flow
just produced by a random number generator. From this simulation, we
extract small-scale Lagrangian flow data, and supply these to a
lattice Boltzmann (LB) model to simulate the flow on scales finer
than the Kolmogorov scale. The LB fluid is then coupled to a
bead-spring chain polymer model, thus providing a fairly realistic
description of both the macromolecule as well as of the surrounding
turbulent flow --- with the big caveat that the input from the large
scales is taken from the behavior of a \emph{Newtonian} fluid, i.~e.
from data that, per construction, \emph{lack} the influence of the
polymeric stress. For this reason, our approach is hampered by a
similar lack of realism as are all the other existing simulation
studies. Therefore, the method should not be viewed as necessarily
superior to previous research, but rather as complementary. Another
caveat is that, even within the restricted setting of the present
approach, it turns out that it is very difficult, if not practically
impossible, within the limitations of existing computational
resources, to match all relevant dimensionless parameters to
experimentally realistic values. This will be the subject of future
publications.

In our opinion, there is no fundamental reason for using a spectral
code for the large scales and LB for the smaller ones. Rather, this
choice was dictated by simple considerations of convenience: The
project is a collaboration between researchers with complementary
expertise in turbulence and polymer solution dynamics, respectively.
In previous studies, turbulent flows had been simulated by the
spectral method~\cite{schumacher:2007}, while the equilibrium dynamics
of polymers in a solvent had been tackled by a hybrid LB-Molecular
Dynamics method~\cite{Ahl:1999,Ahl:2001}, where the coupling between
polymer and solvent was introduced via a Stokes friction coefficient
of the beads. For a detailed description and analysis of this method,
we refer the reader to a recent review, Ref.~\cite{duenwegladd:2008}.
Since the coupling between the two scales was just done via data on a
disk, we saw no reason to not simply continue to run the respective
parts with their corresponding well-tested codes.

In the present paper, we wish to outline (Sec.~\ref{sec:strategy})
the essential features of our multiscale strategy. One important
aspect is that the simulations on the large and small scales
should be mutually consistent, in other words, that the LB solver
on the small scales produces the same flow field as that which
is known from the DNS on the large scales. This will
be the content of Sec.~\ref{sec:validation}, while actual simulation
data with polymer chains, which are currently not yet available
with sufficient accuracy, are deferred to future publications.
Finally, Sec.~\ref{sec:conclus} summarizes our conclusions.

\section{Modelling strategy}
\label{sec:strategy}

The main task is to construct the flow field for the LB-polymer
simulation from the standard DNS of the turbulent flow, where the
latter is done with a strict incompressibility constraint. To this
end, we make use of a passive tracer particle co-moving with the DNS
flow. The tracer particle keeps track of the local environment.
Consider a tracer particle moving with a velocity $\bv(t)$ during
the DNS in a periodic volume of size $L^3$. A cubic sub-volume of size
$l^3$, of order of the Kolmogorov scale, comoving with the tracer, is
defined around the tracer particle, such that the tracer is always located at
the center of mass of the sub-volume. The velocity field in the
sub-volume, $\bu(\br,t)$, is stored during the DNS run. This local flow
field in the sub-volume is to be transplanted into the LB model to
provide the hydrodynamic stress field for the single polymer chain.

As a first step, $\bu(\br,t)$ needs to be Galilei transformed into the
co-moving accelerated frame of the tracer particle, so that there is
no net flux of mass nor of momentum out of the sub-volume, and the
polymer chain, placed at the tracer position, would only interact with
the flow without net translational momentum. The transformed velocity
field, $\bw(\br,t)=\bu(\br,t)-\bv(t)$, is a solution of the same
Navier-Stokes equation with an additional external force density,
which is the mass density $\rho$ times the negative acceleration of
the tracer, $-d \bv/d t$. The resulting local flow field,
$\bw(\br,t)$, is then linearly interpolated and copied into an LB
model in a much finer resolution; the LB flow,
$\tilde{\bw}(\tilde{\br},\tilde{t})$, will be a zoomed-in flow field
of $\bw$ below the scale of the sub-volume $l^3$ of DNS. The LB
resolution needs to be fine enough to model the conformational polymer
degrees of freedom.

Therefore, $\tilde{\bw}(\tilde{\br},\tilde{t})$ is already completely
determined or prescribed by $\bw(\br,t)$. However, in this study, we
do not prescribe $\tilde{\bw}(\tilde{\br},\tilde{t})$ on every mesh
point $\tilde{\br}$ of the LB lattice. Rather, we prescribe
$\tilde{\bw}(\tilde{\br})$ on all sites only at the initial time $t=0$.
At later times, we prescribe it only on the boundaries of the
sub-volume. Since the Reynolds number at the scale of the sub-volume
is small (recall $l \sim \eta$), one must expect that the Navier-Stokes
equation, together with initial and boundary conditions, has one
unique solution. Therefore, the flow generated by the LB solver
with these initial and boundary conditions must be identical
to the input flow. In the following section, we will show that
this is indeed the case, within the intrinsic and unavoidable
errors, which mainly come from the enhanced spatio-temporal
resolution of the LB dynamics, and to a certain extent also
from the fact that the LB simulation works at finite Mach
number, while the DNS code imposes strict incompressibility.

The reason why the information about
$\tilde{\bw}(\tilde{\br},\tilde{t})$ is confined to the boundary is
twofold. On the one hand, we wish to avoid over-constraining the LB
solver as such, which, as discussed, produces a slightly different
solution than the DNS. More importantly, however, we need to allow an
immersed polymer chain to modify the flow field as much as possible.
We thus arrive at a problem that is, at least in principle,
mathematically well-defined: We study the motion of a polymer chain in
a solvent, where the whole system is subject to some time-dependent
boundary conditions, and where in our case these boundary conditions
happen to be derived from the DNS.

Technically, we implemented a rather simplistic approach to enforce
the supplied initial and boundary conditions, by just setting the LB
populations at the pertinent sites to their local equilibrium values.
For the simulations with polymer, we need to avoid that it is
``washed'' out of the simulation box during the course of the
simulation. For this reason, we impose one further constraint, which is
to fix the polymer center of mass at the tracer (the point where
$\bw$ is zero, per construction).

\section{Validation of the strategy}
\label{sec:validation}

In this section, an actual application of the proposed strategy is
introduced and discussed. The DNS of a homogeneous isotropic turbulent
flow was performed in an independent study~\cite{schumacher:2007}. In
what follows, we focus on the parameters (in DNS units) used in the
run number $4$ of Ref.~\cite{schumacher:2007}: $\nu = 1/400$ is the
kinematic viscosity, $\bar{\varepsilon} = 0.1$ is the mean energy
dissipation rate, $R_{\lambda} =
\sqrt{5/(3\bar{\varepsilon}\nu){u_{\mathrm{rms}}}^2} = 65$ is the
Taylor-microscale Reynolds number, where $u_{\mathrm{rms}}$ is the
root mean squared speed of the flow. The large scale Reynolds number,
$Re = u_{rms}L/\nu$, is $2243$, where the box size $L=2\pi$. $M =
1024$ is the number of grid points in each direction of the simulation
box.  The Kolmogorov length~\cite{landau:FM}, $\eta =
{\nu}^{3/4}{\bar{\varepsilon}}^{-1/4}$, becomes $0.0199$ in the length
unit of DNS, which corresponds to $3.26\Delta$, where $\Delta$ is the
grid spacing. The Kolmogorov time,
$\tau_{\eta}=\sqrt{\nu/\bar{\varepsilon}}$, is $0.158$.

\begin{figure}
  \begin{center}
    \includegraphics[clip,width=\figurewidth]{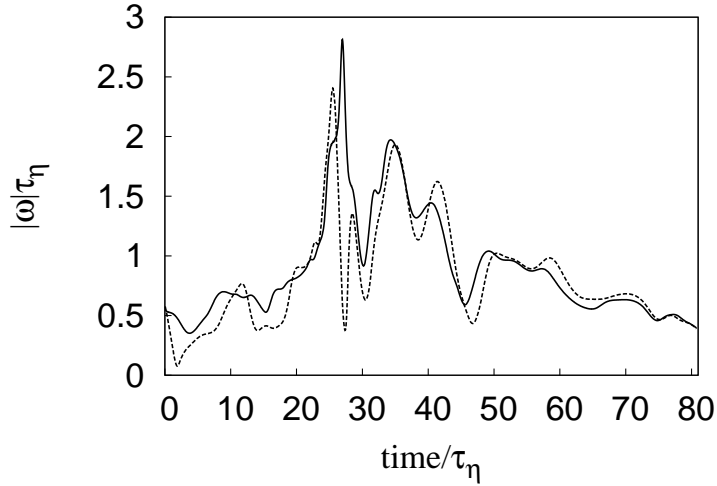}
  \end{center}
  \caption{Magnitude of the vorticity during DNS sampled by a tracer.
  Solid line: The magnitude of the vorticity is averaged over 
  the sub-volume of ${(6\eta)}^3$ around the tracer.
  Dotted line: The magnitude of the vorticity is averaged over 
  an even smaller sub-volume of ${(1.25\eta)}^3$ around the tracer.}
  \label{fig1}
\end{figure}

A set of $100$ tracer particles is used during the DNS. A sub-volume
of ${l^3}=(20\Delta)^3\simeq {(6\eta)^3}$ is defined around each
tracer. The flow field in the sub-volume is monitored at every
$\Delta_{t}=0.05\tau_{\eta}$ for $81\tau_{\eta}$. Figure~\ref{fig1}
shows one example of the flow field sampled by a tracer. The magnitude
of the vorticity vector, $\bm{\omega}(\br,t)=\nabla\times\bw(\br,t)$,
shows strong time dependence during the $81\tau_{\eta}$ of DNS run.
The flow field shows a noticeable spatial inhomogeneity as the two
lines of different control volumes are not identical.

The standard D3Q19 model \cite{duenwegladd:2008} is used for the LB
simulations. In order to parameterize the increased spatial
resolution, we define a ``zooming factor'' via $z=\Delta/a$, where $a$
is the grid spacing of our LB model. The Kolmogorov length in the LB
model is thus $\tilde{\eta}=3.26za$.  Together with the kinematic
viscosity of the LB model $\tilde{\nu}$, the Kolmogorov time in LB
units becomes $\tilde{\tau}_{\eta} = {\tilde{\eta}}^{2}/\tilde{\nu}$.

In prescribing the velocity boundary condition of LB, we match the
reduced velocity between LB and DNS:
\begin{equation}
  \tilde{\bw} \left(\frac{\tilde{\br_{b}}}{\tilde{\eta}},
  \frac{\tilde{t}}{\tilde{\tau}_{\eta}}\right)
  \frac{\tilde{\tau_{\eta}}}{\tilde{\eta}}
  =\bw\left(\frac{\br_{b}}{\eta},\frac{t}{\tau_{\eta}}\right)
  \frac{\tau_{\eta}}{\eta},
\end{equation}
where $\br_{b}$ is the position vector for boundary meshes of the
sampled sub-volume in DNS. $\tilde{\bw}$ has finer resolution than
$\bw$ both in space and time; bilinear interpolation is used
between neighboring grid points and time steps.

\begin{figure}
  \begin{center}
    \parbox{\subboxwidth}{
       \includegraphics[clip,width=\subfigurewidth]{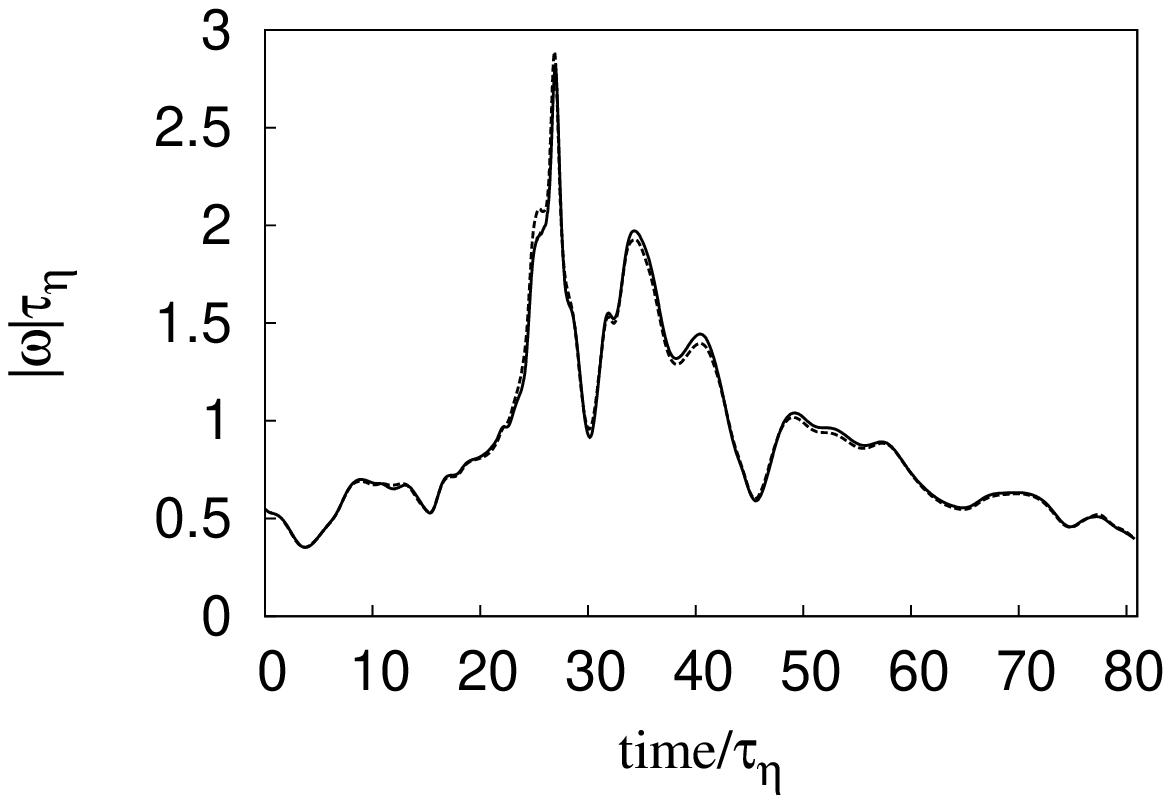}
    }
    \parbox{\subboxwidth}{
       \includegraphics[clip,width=\subfigurewidth]{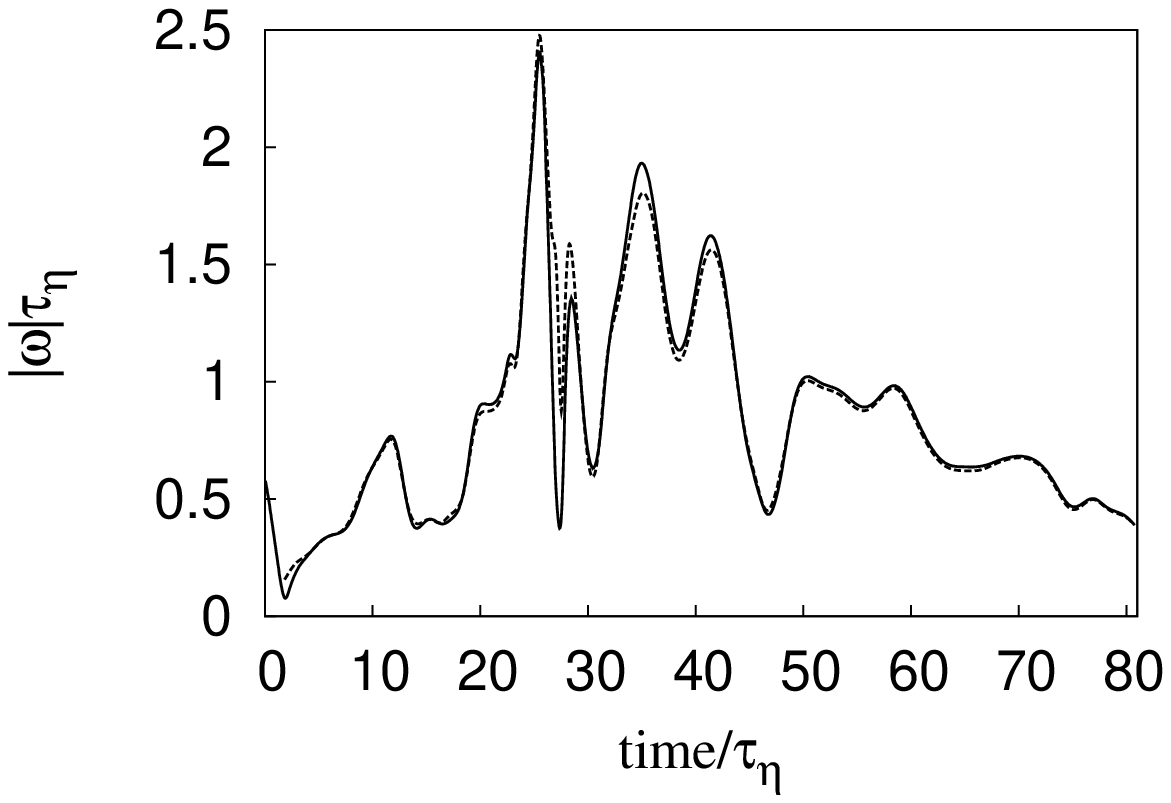}
    }
  \end{center}
  \caption{The magnitude of the vorticity field is compared
  between DNS and LB. The solid lines ($|\omega|\tau_{\eta}$) are
  for DNS, while the dotted lines ($|\tilde{\omega}|\tilde{\tau}_{\eta}$)
  are for LB. (a) The vorticity field is averaged over the total sub-volume of
  ${(6\eta)}^3$. (b) The vorticity field is averaged over a smaller
  sub-volume of ${(1.25\eta)}^3$ around the tracer position.}
  \label{fig2}
\end{figure}

In Fig.~\ref{fig2}, the flow field of LB ($\tilde{\bw}$) is compared
with the flow field of DNS ($\bw$) for the same tracer used for
Fig.~\ref{fig1}. A zoom factor $z=6$ is used: The mesh size of the
sub-volume around the tracer is ${(20\Delta)}^3$ in DNS, while it is
${(120a)}^3$ in the LB model. The magnitude of the vorticity field
over the total sub-volume around the tracer shows a very good
quantitative agreement between DNS and LB, as Fig.~\ref{fig2}(a)
shows. However, we are particularly interested in whether the LB
field, generated only by the boundary condition, would recover the DNS
result in the center region close to the tracer position. In
Fig.~\ref{fig2}(b), the velocity field is compared over a much smaller
region around the center. In general, the LB flow field shows good
agreement with the DNS result even in the region far from the
boundary. However, the LB flow cannot exactly follow sudden changes of
the DNS flow as shown in the time region around $30\tau_{\eta}$.

Averaged over all the flow fields sampled by $100$ tracers, the
difference between the LB flow field and the original DNS flow field
is less than $5\%$ in magnitude compared to the original DNS field. We
believe that the error depends on various parameters of the model,
especially on the Mach number $Ma = \tilde{\eta} /
(\tilde{\tau}_{\eta} c_s)$, where $c_s$ is the speed of sound of the
LB model. The Mach number is $1.49\times 10^{-2}$ for the presented case study.
However, given the fact that for our parameters the
agreement is quite satisfactory, we did not investigate this issue in
a systematic way.

\section{Summary and outlook}
\label{sec:conclus}

A multiscale simulation strategy to increase the resolution of a
local Lagrangian turbulent flow is proposed. The local Lagrangian flow
fields around and below the Kolmogorov scale are sampled by passive
tracers in a standard DNS of turbulent flow. The LB model is used to
regenerate the sampled flow field in a much finer spatio-temporal
resolution.

The strategy proposed in this paper provides a way to generate the
local Lagrangian turbulent flow field for the hybrid Molecular
Dynamics--lattice Boltzmann method of polymer dynamics in
flow~\cite{Ahl:1999,Ahl:2001,duenwegladd:2008}. The whole procedure
provides an unique opportunity to simulate polymer dynamics in a
fairly realistic turbulent flow.

\section*{Acknowledgments}

We thank DEISA, the Distributed European Infrastructure for
Supercomputing Applications, for a generous allocation of computing
resources. JL thanks the Alexander von Humboldt Foundation for
financial support. We thank K. Kremer and E. Bodenschatz for
stimulating discussions.

\end{document}